\documentclass[nopreprintline,3p,a4paper,times]{elsarticle}

\usepackage{graphicx}
\graphicspath{{fig}}

\usepackage{hyperref}

\usepackage{url}
\usepackage{cleveref}
\crefname{figure}{Figure}{Figures}
\crefname{table}{Table}{Tables}
\crefname{section}{Section}{Sections}

\usepackage{framed}
\usepackage{listings,mdframed}
\newcommand{\maru}[1]{\raise0.2ex\hbox{\textcircled{\scriptsize{#1}}}}
\newcommand{\Paragraph}[1]{\textbf{#1.}}

\begin{document}
\begin{frontmatter}

\title{ICCheck: A Portable, Language-Agnostic Tool for Synchronizing Code Clones}

\author[isct]{Motoki Abe}
\author[isct]{Shinpei Hayashi}
\affiliation[isct]{organization={School of Computing, Institute of Science Tokyo},
            addressline={Ookayama 2--12-1}, 
            city={Meguro-ku},
            postcode={152--8550}, 
            state={Tokyo},
            country={Japan}}

\begin{abstract}
Inconsistent modifications to code clones can lead to software defects.
Many approaches exist to support consistent modifications based on clone detection and/or change pattern extraction.
However, no tool currently supports synchronization of code clones across diverse programming languages and development environments.
We propose ICCheck, a tool designed to be language-agnostic and portable across various environments.
By leveraging an existing language-agnostic clone search technique and limiting the tool's external dependency to an existing Git repository, we developed a tool that can assist in synchronizing code clones in diverse environments.
We validated the tool's functionality in multiple open-source repositories, where ICCheck was able to detect overlooked clone modifications in over 30 programming and domain-specific languages and delivered interactive suggestions within a median of 0.27 seconds in editor environments, demonstrating its language independence and responsiveness.
Furthermore, by supporting the Language Server Protocol, we confirmed that ICCheck can be integrated into multiple development environments with minimal effort.
ICCheck is available at \url{https://github.com/salab/iccheck}
\end{abstract}


\end{frontmatter}

\thispagestyle{plain}

\section*{Metadata}

\begin{table}[!h]\centering
  {\footnotesize\begin{tabular}{|l|p{7.75cm}|p{6.7cm}|} \hline
    \textbf{Nr.} & \textbf{Code metadata description} & \\ \hline
    C1 & Current code version & v0.10.0 \\ \hline
    C2 & Permanent link to code/repository used for this code version & \url{https://github.com/salab/iccheck/releases/tag/v0.10.0} \\ \hline
    C3 & Permanent link to Reproducible Capsule & \url{https://doi.org/10.5281/zenodo.15163079} \\ \hline
    C4 & Legal Code License & MIT license \\ \hline
    C5 & Code versioning system used & Git \\ \hline
    C6 & Software code languages, tools, and services used & Go, Kotlin, TypeScript \\ \hline
    C7 & Compilation requirements, operating environments and dependencies & Go v1.23.0+ \\ \hline
    C8 & If available, link to developer documentation/manual & \url{https://github.com/salab/iccheck/blob/main/README.md} \\ \hline
    C9 & Support email for questions & \href{mailto:toki@se.comp.isct.ac.jp}{toki@se.comp.isct.ac.jp}, \href{mailto:hayashi@comp.isct.ac.jp}{hayashi@comp.isct.ac.jp} \\ \hline
  \end{tabular}}
\end{table}

\section{Introduction}\label{s:introduction}

Code clones are identical or similar code fragments~\cite{roy2007survey}.
The presence of code clones is considered a code smell, and refactorings are often performed to eliminate them~\cite{fowler2018refactoring}.
However, some code clones are not suitable for removal due to challenges in refactoring~\cite{mondal-micro-clones}.

Cloned code fragments often require co-evolution.
A set of cloned code fragments is called a clone set~\cite{inoue-code-clone-analysis}.
According to Krinke, in half of the cases where a modification is made to one fragment in a clone set, a consistent modification is also applied to all fragments in the set~\cite{krinke-study-consistent-changes}.
A typical case is that a code fragment with a defect is reused by copy-paste; in this case, the defect of the original fragment may remain in all its cloned fragments, and therefore the same fix may need to be applied to them as well~\cite{roy2007survey}.
Failing to consistently modify code clones in a clone set when co-evolution is required can lead to defects or inconsistent behavior~\cite{juergens-do-code-clones-matter}.
To address this issue, various approaches have been proposed to support consistent synchronizations in a clone set, including clone tracking and change pattern extraction~\cite{toomim-linked-editing, duala-clone-tracker, meng-sydit, ueda-devreplay}.
Here, \emph{clone synchronization} refers to maintaining consistency of modifications across cloned fragments, which differs from \emph{cloning}, the act of creating new duplicates.

However, these techniques and their tool implementations are often designed for specific development environments or programming languages~(see the details in \cref{s:motivation}), so enormous effort is required to adapt these techniques to other environments or languages.
Modern software development involves diverse programming languages~\cite{ieee-spectrum-top-lang} and development environments~\cite{software-ecosystems-github-workflow}, while new languages, environments, and editors are consistently emerging.
Existing approaches for supporting synchronizations cannot be readily applied in such diverse settings.

To address this issue, we propose ICCheck (Inconsistent Change Checker), a tool for assisting with code clone synchronizations that is highly portable and independent of both specific languages and development environments.
In brief, ICCheck works by detecting changes in a Git repository, searching for cloned fragments similar to the modified code, and then identifying and reporting cases where a clone was not consistently updated.
\emph{The impact} of this software can be summarized as follows:
\begin{itemize}
  \item \emph{High portability.}
  ICCheck is applicable to any Git-managed repository that involves code text, regardless of the used programming languages.
  To ensure high portability while minimizing the loss of usability, it can function either as a well-structured command-line interface (CLI) or as a server compliant with the Language Server Protocol (LSP), offering interactive support during code editing.
  Preliminary experiments demonstrated ICCheck's efficiency and portability while maintaining its accuracy (\cref{s:evaluation}).
  To our knowledge, no other tools currently support this combination of features.
  Beyond its strong practical utility, ICCheck is also suitable for empirical studies on clone synchronization across different programming languages.
  \item \emph{Design choices to enable the portability} (\cref{s:motivation,s:technique}).
  To ensure the adaptability across a wide range of use cases and development environments, we carefully selected the underlying clone search algorithm, its usage patterns, and implementation strategies, explicitly documenting them as part of our design choices.
  We believe this provides valuable guidance for future implementations of clone synchronization techniques.
\end{itemize}

The remainder of this paper is structured as follows.
\Cref{s:motivation} introduces a motivating example of code clones for designing ICCheck.
\Cref{s:technique} defines the requirements for ICCheck and presents its design and implementation.
\Cref{s:evaluation} preliminarily evaluates ICCheck.
\Cref{s:conclusion} concludes this paper.

\section{Motivation}\label{s:motivation}

\begin{figure}[tb]\centering
  \includegraphics[width=0.8\textwidth]{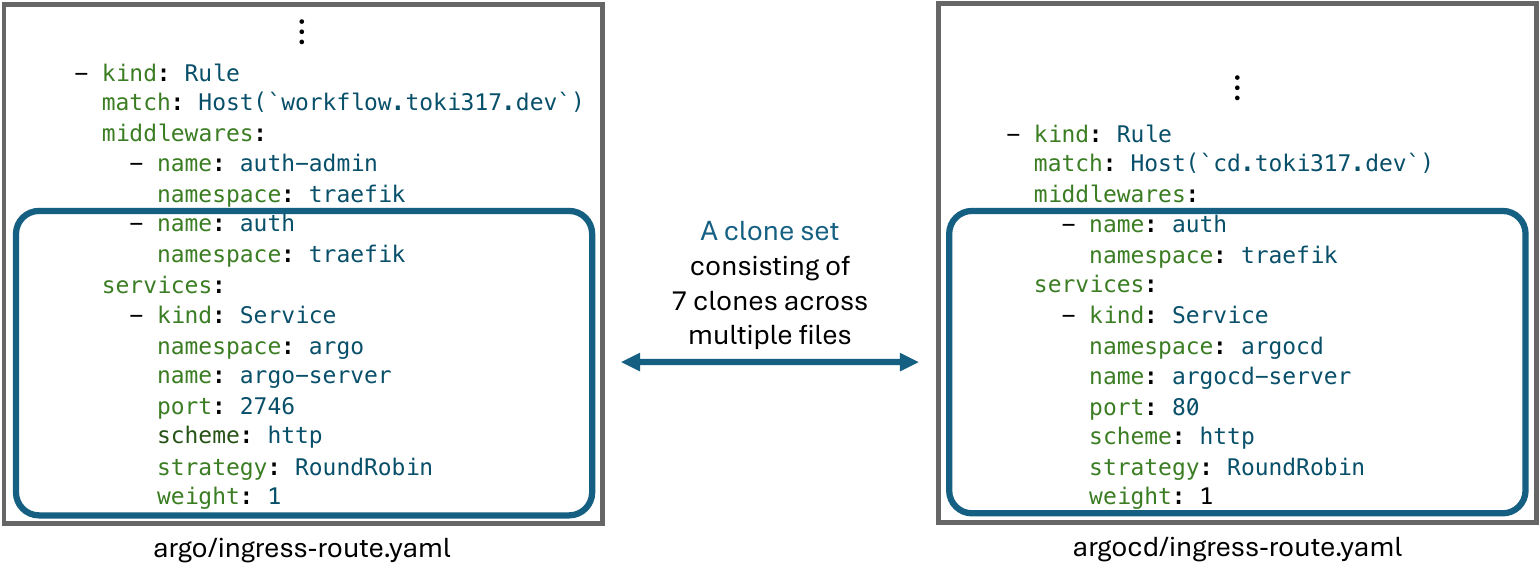}
  \caption{Clone set in YAML files.}\label{f:yaml-code-clone}
\end{figure}

\Paragraph{Code clones in various languages}
Code clones exist not only in general-purpose programming languages such as Java and C but also in domain-specific languages like Dockerfile~\cite{tsuru-dockerfile-clone} and configuration files written in YAML or JSON.
\Cref{f:yaml-code-clone} illustrates an example of code clones found in YAML files\footnote{\url{https://github.com/motoki317/manifest/tree/6e13a913fb8269e0d80d9a6733c3eab7bbcc7cfa}}.
In this case, the file defines Kubernetes object definitions, and whenever modifications are made to a clone instance, all the other instances within the clone set must be updated consistently.
As this example demonstrates, it is not uncommon for code clones to exist across multiple files, even in domain-specific languages with unique syntaxes or in configuration files like YAML and JSON.
As prior studies have shown, code clones exist in a wide range of programming languages~\cite{semura-ccfindersw,zhu-msccd-jss}, highlighting the need for clone management and synchronization techniques applicable across diverse development contexts.
Moreover, consistent modifications are sometimes required for these clones as well.
Therefore, it is essential to support the detection and consistent modification of code clones not only in mainstream programming languages but also in domain-specific languages and data serialization formats used for configurations.
Note that although there is also a demand for detecting clones that span across multiple languages~\cite{prez-crosslang,nafi-CLCDSA,vislavski-LICCA}, our tool does not aim to address the synchronization of such cross-language clones.

\Paragraph{Diversity of use cases and development environments}
In modern software development, code changes are reviewed in various scenarios, including confirming them while editing in an editor, reviewing a sequence of commits in a branch before submitting a pull request, and analyzing them on Continuous Integration/Delivery~(CI/CD) platforms.
Developers use a variety of tools for editing and validating source code, ranging from simpler code editors to advanced text editors and integrated development environments (IDEs) such as Visual Studio Code (VSCode) and IntelliJ IDEA.
Additionally, numerous CI/CD platforms, such as GitHub Actions, are used for validating and revising source code.
To effectively support simultaneous modifications to code clones in real-world software development, a highly portable tool is required: one that integrates seamlessly with various text editors and can also be executed in CI/CD environments.

Based on these motivations, the following key aspects are crucial for supporting consistent modifications of code clones in practical software development:
  1) the support for simultaneous modification of code clones in minor domain-specific languages and configuration file formats, and
  2) high portability, enabling usage across diverse use cases and development environments.

\begin{table}[tb]\centering
  \caption{Comparison of approaches for supporting clone synchronization}\label{t:related-works}
  {\footnotesize\tabcolsep=3pt\begin{tabular}{lllll} \hline
    Name & Languages & Environments & Tool availability \\ \hline
    Simultaneous Editing~\cite{miller2001interactive} & Java, HTML & LAPIS & Available~(\url{https://groups.csail.mit.edu/graphics/lapis/}) \\
    Linked Editing~\cite{toomim-linked-editing} & Java & XEmacs & Not specified \\
    CloneTracker~\cite{duala-clone-tracker,duala-clone-tracker-tool} & Java & Eclipse & Available~(\url{https://www.cs.mcgill.ca/~swevo/clonetracker/}) \\
    CReN~\cite{jablonski-cren} & Java & Eclipse & Link broken \\
    JSync~\cite{nguyen-jsync} & Java & Eclipse & Not specified \\
    CCSync~\cite{cheng-ccsync} & Java & Not specified & Not specified \\
    CCDemon~\cite{lin-ccdemon} & Java & Eclipse & Available~(\url{https://github.com/llmhyy/ccdemon}) \\
    Clone Notifier~\cite{clone-notifier} & Covered by 3 detectors & Custom GUI & Available~(\url{https://github.com/s-tokui/CloneNotifier}) \\
    CLIONE~\cite{clione} & Java, Kotlin, Python, C++ & CI (GitHub App) & Available~(\url{https://github.com/T45K/CLIONE}) \\
    CloneTracker (Commercial)~\cite{fixstars-clonetracker} & 10 languages & Custom GUI & Available~(\url{https://clonetracker.com/en/}) \\
    ICCheck (our tool) & Language-agnostic & CLI, CI, LSP clients & Available~(\url{https://github.com/salab/iccheck}) \\ \hline
  \end{tabular}}
\end{table}
Despite the advancements in code clone research and tools, to the best of our knowledge, no existing tool meets these requirements.
\Cref{t:related-works} provides a comparison of existing approaches for supporting simultaneous modifications based on code clones, including those covered in a recent survey of clone tracking techniques~\cite{mondal-survey-clone}, revealing the supported languages and environments, and the tool availability.
While various techniques and tools have been proposed, most of them are limited to Java and are implemented for specific IDEs, such as Eclipse or custom editors.
Furthermore, some tools are not publicly available.

\section{ICCheck in a Nutshell}\label{s:technique}

\subsection{Design}\label{s:technique-design}

The key design choices in the development of ICCheck are outlined below.

\Paragraph{(1) Integrating language-agnostic clone search technique}
ICCheck detects source code changes made by developers, identifies code clones containing similar code snippets with the changed ones, and notifies the developer of their presence.
For this purpose, clone search, where a set of modified code fragments serves as a query to find similar code fragments, is more suitable than full-scale clone detection, which identifies duplicated sections across an entire codebase.
However, a language-agnostic clone search method is required to ensure adaptability to any programming language without additional effort.
In particular, detecting \emph{micro-clones}, which are code clones shorter than five lines of code, is crucial~\cite{mondal-micro-clones}.
To achieve this, ICCheck employs FLeCCS~\cite{mondal-fleccs}, a language-independent technique that identifies clones based on textual similarity with the given query code fragment.
FLeCCS compares all files in a target project using a sliding window approach.
It calculates bigram sets at the character level for each line in the query and the target code snippet, then it determines similarity using the weighted average of Dice coefficients with weighting based on line length.
By default, a pair of code fragments with a similarity score of 0.7 or higher is regarded as a code clone, following the default setting of FLeCCS; however, this threshold can be changed through a user-specified option.

\Paragraph{(2) Adherence to Git-based code change processes}
To efficiently apply ICCheck across various code change scenarios while keeping implementation costs low, it leverages Git, a widely used version control system, to track changes.
The relevant use cases discussed in \cref{s:motivation} can be recognized as specific changes within the version control context.
By integrating with Git, ICCheck seamlessly incorporates the support of clone synchronization into the most common software development workflows today.
ICCheck identifies a changeset by specifying commit IDs, branch names, or the working tree before committing.
It searches for related code clones within the project using the changeset as a query and notifies the developer of their presence.
In addition, ICCheck operates solely on a Git repository, without requiring any precomputed clone database or additional intermediate artifacts.
By ensuring that it performs cleanly without leaving any analysis artifacts, ICCheck remains simple to use and easy to integrate into existing workflows, thereby reducing the barrier to adoption.
We therefore intentionally chose not to adopt more computationally intensive, history-aware clone detection approaches that verify the co-evolution of cloned fragments across versions, which are typically heavy and expensive as they rely on precomputed clone databases or repeated cross-version analyses.

\Paragraph{(3) Providing CLI and Language Server}
To interactively support clone synchronization in diverse environments, ICCheck offers both a Command Line Interface (CLI) and integration with the Language Server Protocol (LSP)~\cite{microsoft-lsp}, which is supported by many modern development environments.
Clone search can be executed via the CLI by specifying a Git repository and changeset, and the results can be in human-readable text or machine-friendly JSON formats.
By leveraging CLI output and exit codes, ICCheck can be integrated with CI/CD workflows such as GitHub Actions and commit-hooks.
Additionally, by implementing a Language Server, ICCheck enables integration with multiple editors, including VSCode and IntelliJ, allowing interactive support during code editing to detect potential overlooked changes with reasonable implementation cost.

\begin{figure}[tb]\centering
  \includegraphics[width=0.75\textwidth]{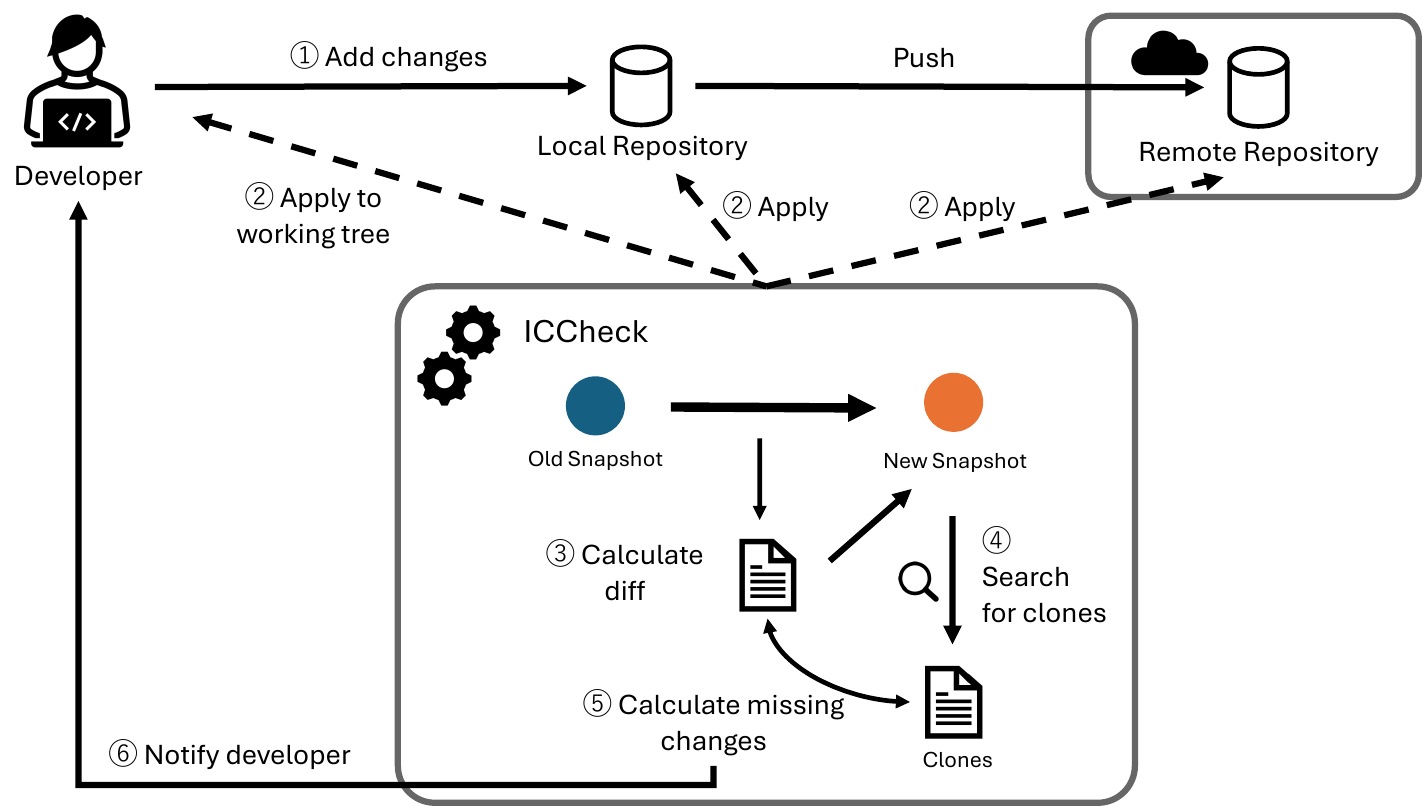}
  \caption{Workflow of ICCheck.}\label{f:iccheck-overview}
\end{figure}

The workflow of ICCheck is summarized in \cref{f:iccheck-overview}.
The developer adds changes to source code in a Git repository (\maru{1}) and runs ICCheck, which conceptually takes two snapshots as input to define the change range (\maru{2}).
Then, ICCheck computes code differences between the two snapshots (\maru{3}).
Using the modified code as a query, ICCheck searches for code clones in the new snapshot (\maru{4}).
It detects overlooked change opportunities by comparing the identified clones with the modified code (\maru{5}).
Any unchanged clone instances are considered overlooked, and they are notified to the developer (\maru{6}).

\subsection{Implementation}\label{s:implementation}

To distribute both the CLI and Language Server as a single executable file while supporting multiple operating systems (Windows, macOS, and Linux) and architectures (amd64 and arm64), ICCheck was implemented from scratch using Golang.
The underlying clone search algorithm, FLeCCS, was reimplemented in Golang.
Although it was originally implemented as a Java GUI application, we did not use it for the ease of implementing the integration with Git.
The latest version (v0.9.0 at the time of writing) is available on the GitHub releases page with its executables.

\begin{figure}[tb]\centering
  \includegraphics[width=0.6\textwidth]{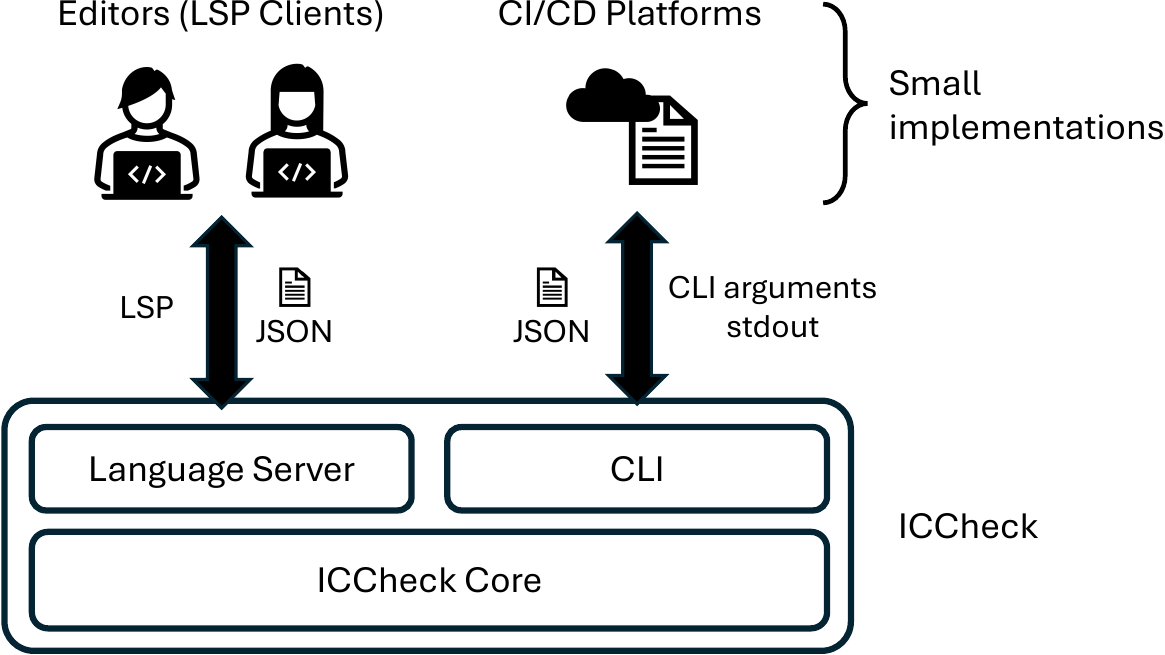}
  \caption{Architecture of ICCheck.}\label{f:iccheck-module-design}
\end{figure}

The architecture of ICCheck is illustrated in \cref{f:iccheck-module-design}.
The CLI and Language Server modules handle external input and output.
The Core module performs clone search and computes overlooked changes.
This modular design keeps the implementation simple and maintainable.
By providing a Language Server, ICCheck enables integration with LSP clients.
Additionally, the CLI module's output can be structured in JSON format, making it easy to process programmatically.
This simplifies CI/CD setup, allowing users to format the output for GitHub Actions with simple one-line shell commands.

ICCheck aims to provide suggestions within 1 second in an editor environment~\cite{card-information-visualizer} and within 1 minute via the CLI.
To achieve this, several optimizations were implemented such as parallel clone search using goroutines.
When running as a Language Server, incremental clone search for code differences is performed, and ICCheck sleeps if the execution time exceeds a threshold and resumes when the developer's keystrokes pause to mitigate the runtime load in the developer's local environment.

\subsubsection{The CLI}

\begin{figure}[tb]\centering
\begin{framed}\begin{lstlisting}
$ iccheck
2024/11/26 22:43:13 INFO 22 change chunk(s) within 3 file(s) found. 
    from="HEAD (d46bf7e87fb62877f0052534659589ecc4c8aa41)" to=WORKTREE
2024/11/26 22:43:13 INFO 5 clone(s) are likely missing consistent change.

Clone set #0 - 5 out of 6 clones are likely missing consistent change(s).
  Missing changes (5):
    pkg/lsp/handler.go:74 (L74-L74)
    pkg/lsp/handler.go:93 (L93-L93)
    pkg/lsp/handler.go:112 (L112-L112)
    pkg/lsp/handler.go:147 (L147-L147)
    pkg/lsp/handler.go:52 (L52-L52)
  Changed clones (1):
    pkg/lsp/handler.go:167 (L167-L167)
\end{lstlisting}\end{framed}
  \caption{Usage of ICCheck CLI.}\label{f:cli}
\end{figure}

ICCheck is distributed as a single executable file with basic usage instructions documented in its README in GitHub.
For example, when the CLI is triggered at a Git-managed directory with a modified working tree, it outputs related code clones and missing changes as shown in \cref{f:cli}.
In addition to the basic workflow described in \cref{s:technique-design}, the tool first checks whether the current directory is Git-managed and, if so, sets the detected repository as the search target.
The target repository can also be changed using a CLI option.
Although the before- and after-change snapshots can be explicitly specified via CLI options, if they are not provided, the tool automatically determines the most appropriate comparison based on the state of the Git repository.
In this example, since the working tree was modified but not yet committed, the comparison was made between the HEAD commit and the working tree.
Finally, the identified suspicious clones are displayed as ``missing changes''.

Executing \texttt{iccheck {-}{-}help} provides a list of CLI options and their descriptions.
The \texttt{{-}{-}repo}, \texttt{{-}{-}from}, and \texttt{{-}{-}to} options specify the Git repository and the snapshots before and after changes, respectively.
If \texttt{{-}{-}from} is not specified, ICCheck automatically uses the parent of the commit specified by \texttt{{-}{-}to} as the snapshot before changes.
The \texttt{{-}{-}ignore} option allows filtering of the reported missing changes, whereas the \texttt{{-}{-}include} option limits the paths to be analyzed.
The \texttt{{-}{-}format} option specifies the output format with \texttt{{-}{-}format json} being particularly useful for CI/CD integration.

Output clone instances are displayed in the order they are detected, i.e., according to the order of files processed, since ICCheck does not employ a specific ranking algorithm.
Users who prefer a particular ordering can still implement their own ranking criteria, such as those proposed by Hamid et al.~\cite{hamid-fleccs-ranking}, as a post-processor of the JSON output.

\subsubsection{The Language Server}

ICCheck integrates with any LSP-compatible editor via its Language Server.
Some LSP-compatible editors can utilize ICCheck without requiring special plugins simply by configuring the Language Server binary path.
For environments that this simple configuration is not available, we implemented simple adapter plugins for VSCode\footnote{\url{https://marketplace.visualstudio.com/items?itemName=motoki317.iccheck}} and IntelliJ IDEA\footnote{\url{https://plugins.jetbrains.com/plugin/24779-iccheck--inconsistency-check}} to make ICCheck runnable easily.

\begin{figure}[tb]\centering
  {\footnotesize\begin{tabular}{cc}
    \includegraphics[width=0.4\textwidth]{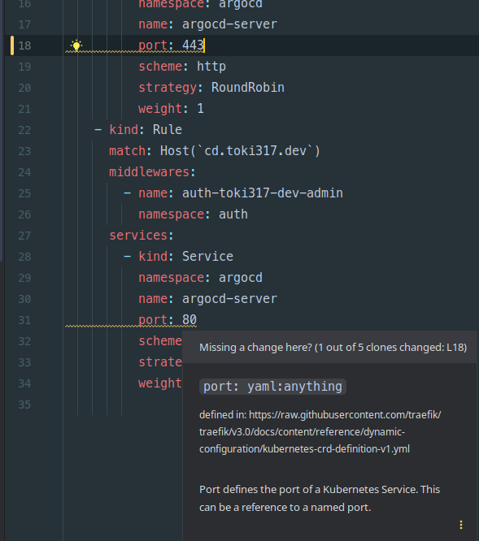} &
    \includegraphics[width=0.4785\textwidth]{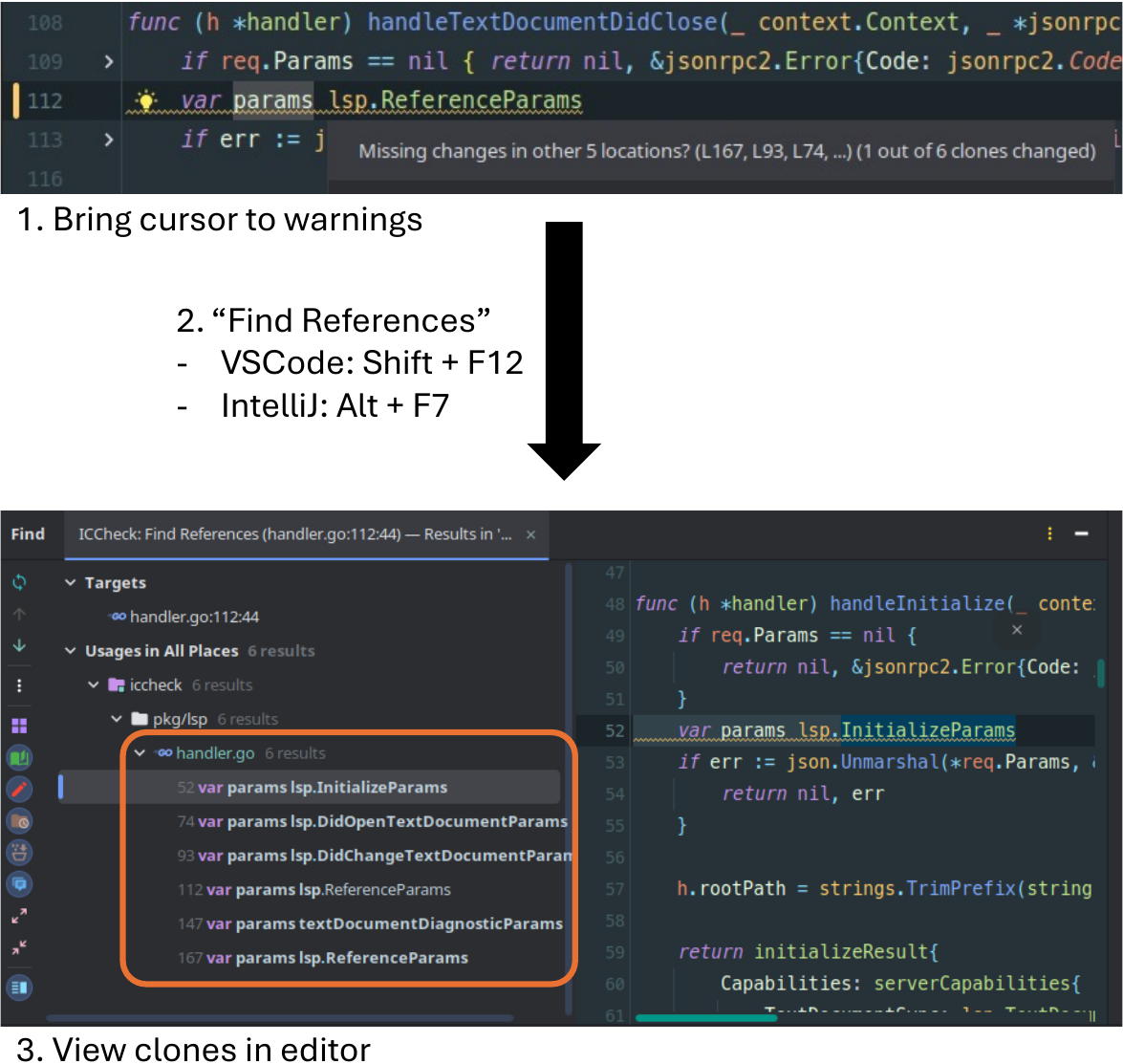} \\
    (a) Highlighting clone locations. &
    (b) Displaying list of clone locations.
  \end{tabular}}
  \caption{Supporting clone synchronization via ICCheck Language Server.}\label{f:lsp}
\end{figure}

When changes are made to a file within a Git repository, the Language Server detects the changes and searches for corresponding code clones.
For example, in the YAML file shown in \cref{f:lsp}(a), modifying the \texttt{port} field on Line~18 triggers a warning suggesting a change to the unchanged micro-clone at Line~31.
Additionally, as illustrated in \cref{f:lsp}(b), hovering the cursor over a warning and using the ``Find References'' feature allows quick navigation to all corresponding clone locations, without leaving the code editor.
These warnings can serve as mild, real-time hints rather than strict alarms.
In this sense, they help developers remain aware of potentially relevant clones while coding, without interrupting their workflow or requiring exhaustive inspection of all reported fragments.

\subsubsection{Implementation Strategies and Optimizations}

To improve runtime performance and scalability, ICCheck incorporates several design and implementation optimizations, as summarized below.
\begin{itemize}
  \item
  During detection, temporary data used for searching \emph{within each file} are discarded immediately after that file's search completes.
  As a result, the only data accumulated in memory across the execution are the detected missing changes to be reported, enabling detection without sustained memory pressure.
  For example, when applied to the latest commit of the Linux kernel repository (41.3M lines of code)\footnote{\url{https://github.com/torvalds/linux/commit/9b332ce}}, the analysis completed using only 3.7 GB of memory.
  \item
  We have ensured that the implementation of the hot path in the FLeCCS algorithm is efficient enough.  
  In particular, the bigram set of each line is represented as a sorted integer array, and intersections are counted via a two-pointer scan, which significantly reduces computational overhead compared with set-based approaches.
  \item
  We extended the \textit{go-git} library in Go to parallelize the loading of Git tree structures, accelerating traversal and metadata reads on large repositories.
  \item
  In LSP mode, detection can be triggered on every edit.
  To avoid redundant work, per-file results from past queries are cached and reused when a file remains unchanged under the current edit sequence; only invalidated files are recomputed.
  Additionally, when ICCheck runs longer, the process inserts short sleeps to avoid monopolizing CPU and to prevent disruption of developer activities.
\end{itemize}

\subsection{Evolution of ICCheck}

\begin{table}[tb]\centering
  \caption{Releases of ICCheck}\label{t:evolution}
  {\footnotesize\begin{tabular}{lll} \hline
    Release Date & Version & Summary \\ \hline
    2024--05--29 & v0.1.0 & Initial release \\
    2024--06--20 & v0.3.0 & LSP functionality \\
    2024--10--21 & v0.5.0 & LSP performance improvements \\
    2024--11--17 & v0.6.0 & Displaying clone locations in LSP \\
    2024--11--26 & v0.7.0 & Suggestion exclusion feature, CLI option reorganization \\
    2024--12--05 & v0.7.1--0.7.6 & Performance optimizations \\
    2025--01--12 & v0.8.0--0.8.2 & Performance optimizations and interface improvements \\
    2025--01--26 & v0.9.0 & CI workflow and minor enhancements \\
    2025--10--14 & v0.10.0 (Latest) & CLI option for specifying the path to be included \\ \hline
  \end{tabular}}
\end{table}

Since its initial release (v0.1.0), ICCheck has undergone continuous development for approximately eight months.
A total of 34 releases have been made, focusing on performance improvements, bug fixes, and feature enhancements, contributing to its maturity as a tool.
\cref{t:evolution} summarizes the key releases in the evolution of ICCheck.
For example, the suggestion exclusion setting was introduced in v0.7.0.
Previous research~\cite{zhang-predicting-clone} has shown that not all fragments in a clone set always need to be evolved for the same reason.
Through our usage experiences, we found that automatically generated files and \texttt{import} statements were frequently reported as false positives.
To address this issue, a feature allowing exclusions via CLI options and/or configuration files was implemented.

Furthermore, improvements were made based on over two months of practical use by multiple practitioners and the feedback obtained.
They provided positive feedback on ICCheck's ability to identify missing changes in YAML configurations and eRuby templates during new feature development involving copy-pasting.
The capability to extend beyond standard programming languages makes it particularly useful for maintaining consistency in diverse file types.

\section{Preliminary Evaluation}\label{s:evaluation}

We quantitatively evaluated whether ICCheck meets the required criteria from three perspectives: execution time, portability to various languages, and accuracy.
In the following experiments, we used ICCheck v0.8.2, which was the latest version available at the time of the evaluation\footnote{%
  The updates from the version evaluated in this study (v0.8.2) to the latest version (v0.10.0) include a minor improvement to the detection behavior that may affect the results of the experiments on open-source repositories (Sections 4.1, 4.2, and 4.3), although the impact is expected to be small and unlikely to affect the conclusions in these subsections.%
}, on Ubuntu 24.10 (WSL2) running on an i9-12900K (16 cores, 24 threads) with 16 GB RAM.
Several scripts for reproducing the experimental results are provided in the repository\footnote{\url{https://github.com/salab/iccheck/tree/main/docs/evaluations}}.

\begin{figure}[tb]\centering
  \begin{minipage}[b]{0.53\textwidth}\centering
    \includegraphics[height=0.22\textwidth]{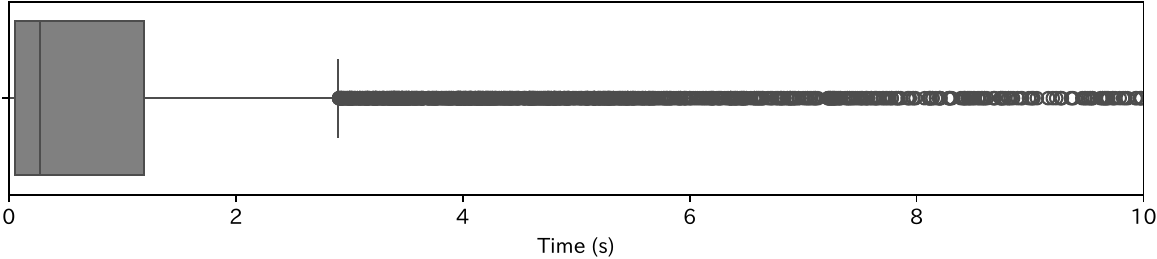}
    \caption{Execution time of ICCheck (seconds, $n={}$9,314).}\label{f:rq1-run-times}
  \end{minipage}
  \begin{minipage}[b]{0.45\textwidth}\centering
    \includegraphics[height=0.225\textwidth]{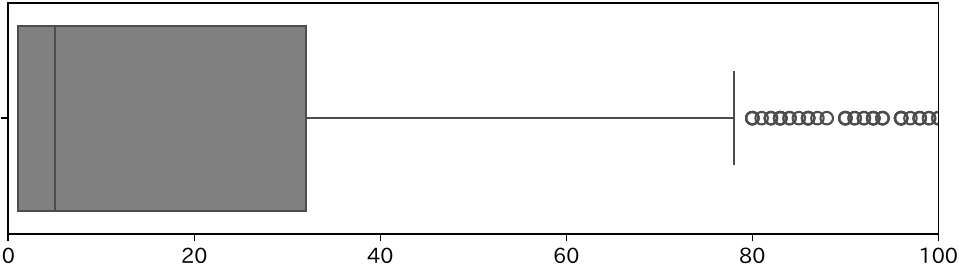}\vspace{0.25cm}
    \caption{Number of detected missing changes ($n={}$2,677).}\label{f:rq2-detected-changes}
  \end{minipage}
\end{figure}

\subsection{Execution Time}

Based on the GitHub Ranking as of August 7, 2024~\cite{github-evanli-ranking}, we selected the top three repositories from each of the 34 language categories, totaling 102 repositories.
Although this star-based selection approach is not highly reliable and may include repositories that are not necessarily well-engineered projects~\cite{repo-reapers}, we prioritized its simplicity and ease of obtaining key data under the constraint of selecting repositories across a wide variety of programming languages.
For each repository, we analyzed the latest 100 commits on the default branch.
However, commits with more than 25 modified files were excluded, resulting in a total of 9,368 commits being analyzed for missing changes.
This threshold of commit size was heuristically chosen to exclude large-scale commits, which are likely to involve perfective maintenance operations~\cite{large-commits}, such as refactorings or search-and-replace edits, rather than ordinary interactive coding activities, as such large commits may introduce noise into the analysis.
The timeout for ICCheck was set to 60 seconds.

Among the 9,368 commits, 9,314 (99.4\%) completed the detection without timing out.
The distribution of their execution times are shown in \cref{f:rq1-run-times}.
For easier viewing, the x-axis is cropped to 10 seconds.
The median execution time was 0.27 seconds with an average of 2.12 seconds.
75\% of the cases completed within 2 seconds.
However, in some cases with large changesets or substantial repository sizes, detection took longer.
For example, the detection in \texttt{torvalds/linux} took at least 14 seconds per commit.

In conclusion, \textit{ICCheck CLI can provide suggestions with a practical response time for most repositories.}

\subsection{Portability to Various Languages}

\begin{figure}[tb]\centering
  \includegraphics[width=0.6\textwidth]{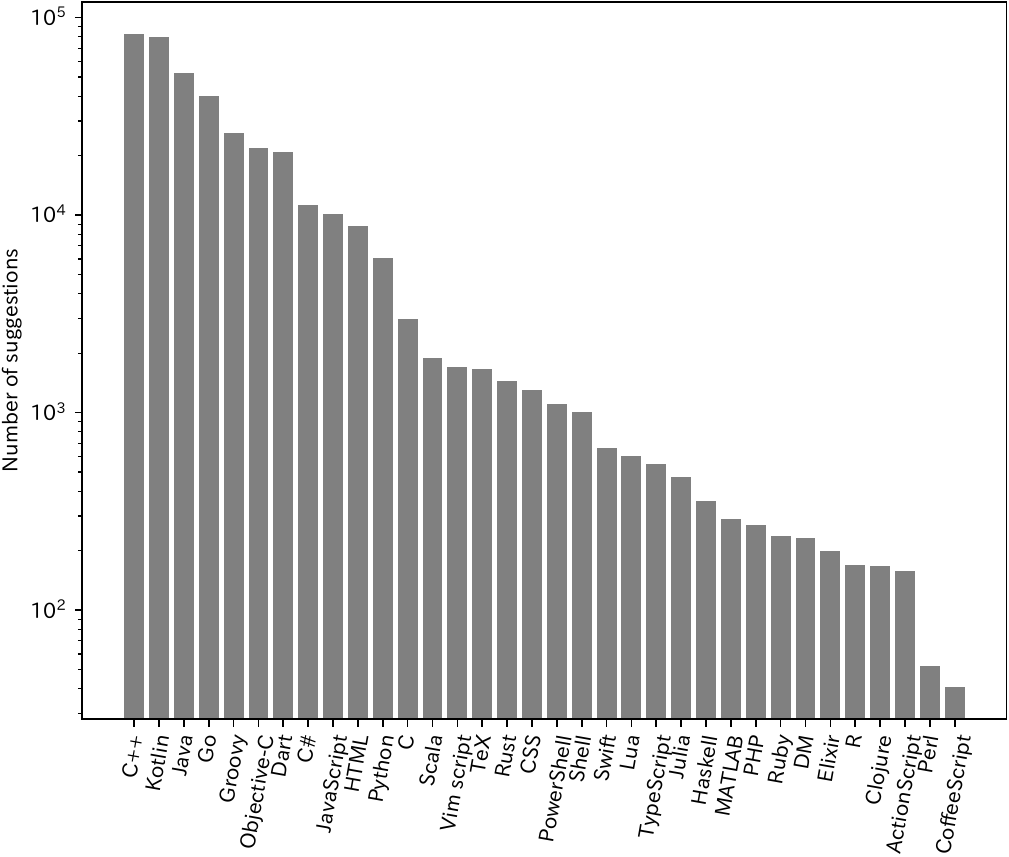}
  \caption{Number of suggestions per programming language.}\label{f:rq2-lang-info}
\end{figure}

Analyzing the results from the 9,314 commits that did not time out, missing changes were detected in 2,677 commits (28.7\%).
As shown in \cref{f:rq2-detected-changes}, the median number of detected missing changes per commit was 5, the average was 225, the maximum was 53,005, and the total was 602,416.
Due to the large maximum value, the axis in the figure is cropped at 100.
As shown in \cref{f:rq2-lang-info}, change suggestions were confirmed for all 34 programming languages used in the study.
Note that these repositories also included data formats such as JSON, which were not included in the selected 34 languages, and suggestions for them were also observed.

In conclusion, \textit{ICCheck is capable of assisting with clone synchronization across multiple languages.}

\subsection{Accuracy for Open-Source Repositories}

From the suggestions obtained in the previous evaluation on the portability, we randomly sampled 100 clone sets and manually validated them.
As a result, 63 out of 100 were correct code clones, of which 49 required consistent synchronization for all their clone instances, and 14 were not.
Meanwhile, 37 were non-clones, i.e., false positives, consisting of unrelated similar code or similar lines formed by auto-generated files, resulting in incorrect suggestions.
The correct clone set suggestions included languages such as TeX, CSS, JavaScript, TypeScript, Dart, Vim script, PowerShell, Python, Julia, Lua, and Go.
Moreover, cross-language change suggestions were also confirmed, such as when code snippets in Markdown triggered suggestions for TeX snippets.

In conclusion, \textit{ICCheck results included meaningful clone suggestions in almost half of the cases, covering multiple languages.}

\subsection{Accuracy for the CBCD Dataset}

We used the dataset of the CBCD approach~\cite{li-cbcd}, which involves 53 C language clone sets containing bugs.
Note that we also referred to the NCDSearch dataset~\cite{ishio-ncdsearch} for the location information of the bugs in the CBCD dataset.
Out of these 53 cases, we excluded 15 clone sets that did not exist within the same snapshot and evaluated the remaining 38 cases.
The average precision and recall per bug were 0.281 and 0.289, respectively.
While there are 69 correct bug locations to predict across the entire dataset, ICCheck produced 23 suggestions, 16 of which were correct.
Thus, the suggestion-based precision was 0.696, and recall was 0.232.
The baseline approach, FLeCCS~\cite{mondal-fleccs}, reports a precision and recall of around 0.6 and 0.5, respectively.
Although the obtained precision, including that of the baseline approach, is not particularly high, this may partly stem from our design choice of a lightweight, on-demand approach that analyzes a single snapshot rather than maintaining historical clone data across versions.
Despite this trade-off, ICCheck's suggestion mechanism provides practically useful interactive assistance by identifying potentially overlooked clone modifications in practical development scenarios.

In conclusion, although the evaluation targets differ, \textit{the precision of ICCheck was comparable to the evaluation results of FLeCCS.}

\section{Conclusion}\label{s:conclusion}

By utilizing a language-independent clone search technique, limiting external dependencies to Git, and adopting a CLI-based input/output approach, we designed and prototyped ICCheck, a highly portable tool for supporting code clone synchronization.
Furthermore, by incorporating Language Server functionality, we demonstrated that ICCheck can be easily integrated with various editors.
We also provided a quantitative evaluation showing that ICCheck can support clone synchronization in multiple languages within a practical execution time and that its suggestion accuracy is reasonable.

As revealed in the execution time evaluation, execution time increases in proportion to the volume of code differences.
This is because ICCheck treats each fragment of a code difference as a query and performs clone search across the entire snapshot referenced by the target commit.
Optimizing the suggestion algorithm may help improve execution time.
Another direction for improvement is to conduct a sensitivity analysis to empirically identify a more appropriate default similarity threshold that contributes to overall accuracy, as well as a comprehensive comparative evaluation of ICCheck against existing clone-synchronization approaches to assess their relative effectiveness and efficiency in detecting overlooked clone modifications across diverse programming languages.
Other future work includes improving the usability and stability of ICCheck by continuing and expanding its use in practical environments, which will help us fix bugs and add new features.
After these improvements, conducting more formal user studies and surveys to systematically evaluate the practical usefulness and usability of ICCheck in real development environments will be valuable.

\section*{Acknowledgments}
This work was partly supported by JSPS KAKENHI (JP23K24823, JP25K03102, JP25H01125, JP24H00692, and JP21KK0179).
We would like to thank Dr.\ Michael J.\ Decker for his comments on the earlier version of this manuscript.

\section*{Declaration of generative AI and AI-assisted technologies in the writing process}
During the preparation of this work the authors used ChatGPT in order to improve readability and language of the work.
After using this tool/service, the authors reviewed and edited the content as needed and take full responsibility for the content of the publication.

\bibliographystyle{elsarticle-num}
\bibliography{references}

\end{document}